\documentclass[aps,pre,reprint,groupedaddress]{revtex4-2}
\usepackage{graphicx}
\usepackage{amsmath}
\usepackage{amsfonts}
\usepackage{subfigure}
\usepackage{color}

\begin{document}


\title{Typicality of thermal states in isolated quantum systems \\ corresponds to ubiquity of global minima in wide artificial neural networks}    


\author{Takaaki Monnai}
\affiliation{Department of Science and Technology, Seikei University, Tokyo, 180-8633, Japan}


\date{\today}

\begin{abstract}
The Neural Tangent Kernel theory theoretically guarantees the existence of global minima of the cost functional in the neighborhood of an arbitrary random initialized parameters in wide artificial neural networks. 
In this paper, we show that the ubiquity of the global minima directly corresponds to the typicality of pure thermal states in isolated quantum systems by identifying a common underlying mechanism characterized by the restriction to a few observables and the role of a Wishart-type matrix. 
Moreover, we demonstrate that the increase in distinguishability of the reduced density matrices of typical pure states with subsystem size corresponds to the double descent phenomenon observed by varying the width of layers in finite-width artificial neural networks. Thereby, the threshold for the reduced state become thermal is determined by essentially the same condition as the fitting threshold. 
In this manner, we reveal a structural correspondence between thermalization in isolated quantum systems and wide neural network.
\end{abstract}


\maketitle

\section{Introduction} 
In recent years, machine learning has achieved remarkable progress.
Powerful AI systems based on Artificial Neural Networks (ANNs) have been realized through overparameterization. The theoretical foundation underlying this success has been significantly advanced by the Neural Tangent Kernel (NTK), which provides important insights into the role of overparameterization in the function representation capability of ANNs\cite{Jakot1}.
The NTK theory describes how an ANN with sufficiently wide layers determines an unknown function from given input-output data points.
In this regime, training error during kernel descent can be driven close to zero while remaining within a small neighborhood of the randomly initialized parameters.
From the physics side, a correspondence between the behavior of entanglement entropy during black hole evaporation and the double descent phenomenon in quantum machine learning has been pointed out, attracting significant interest\cite{blackhole1}.
However, since black hole evaporation is a nontrivial but rather special astrophysical phenomenon, it is desirable to clarify a more universal correspondence between physics and machine learning.

As a first step, in this paper we focus on the universal phenomenon of thermalization and reveal a structural correspondence between the typicality of pure thermal states\cite{Lebowitz1,Popesucu1,Sugita1,Reimann1} and the ubiquity of the global minima of the cost functional in the overparameterized regime of NTK in terms of the restriction of observables and overparametrization\cite{Jakot1}. 
We also show a further correspondence between the system size dependence of the distinguishability of the reduced density matrices of typical pure states and the double descent phenomenon in artificial neural networks, where the generalization error decreases again as the number of parameters increases.
Thermalization in isolated quantum systems is manifestly more general than black hole evaporation, and the correspondence with NTK is underpinned by a shared framework characterized by a restriction to a small number of observables, overparameterized degrees of freedom, and the central role of Wishart-type matrix. 
Therefore, we establish a structural correspondence between key concepts in deep learning such as the ubiquity of global minima, double descent, and the fitting threshold and their counterparts in thermalization. 
The purpose of this paper is to clarify a nontrivial structural correspondence and to offer a first step toward a unified theoretical perspective linking thermalization in isolated quantum systems and deep machine learning.

Before outlining the structure of this paper, we briefly recapitulate the thermalization in isolated quantum systems and NTK framework.

The relaxation processes in the unitary time evolution of isolated quantum systems have been extensively studied both numerically\cite{Shankar1,Rigol1} and experimentally\cite{Schmiedtmayer1,Trozky1}, particularly in cold atomic systems.
From a theoretical point of view, thermalization has been investigated based on spectral fluctuations, that is, intrinsic thermal properties of quantum systems such as typicality and quantum ergodic theory\cite{vonNeumann1,Lebowitz2}.  
In both cases, thermalization in isolated quantum systems is facilitated by the high dimensionality of the Hilbert space. 

On the other hand, in machine learning theory, NTK describes the behavior of artificial neural networks in the infinite-width limit.
In this regime, the network's training dynamics are governed by a fixed kernel, allowing the network to effectively perform kernel regression that interpolates the target function based only on its values at the input data points.

This paper is organized as follows. 
In Sec. II, we explore the correspondence between typicality of pure thermal thermal states and ubiquity of global minimum. 
In Sec. III, we also investigate the correspondence between the system size dependence of the distinguishability of the reduced density matrix and the double descent phenomenon. 
Sec. IV is devoted to a summary. 
\begin{table*}[t]
\caption{Correspondence between thermalization in isolated quantum systems and function estimation in artificial neural network.}
\centering
\begin{tabular}{|l|l|}
\hline
\textbf{Thermalization in isolated quantum systems} & \textbf{Function estimation in artificial neural network} \\
\hline
High dimensionality of the energy shell & Overparameterization \\
Observable $\hat{O}$ & Function $f(x,\theta)$ \\
Typicality of pure thermal states & Ubiquity of global minima in NTK \\
System size dependence of entanglement entropy & Double descent phenomenon \\
Maximum entanglement entropy & Fitting threshold \\
\hline
\end{tabular}
\label{correspondence1}
\end{table*}
\section{Typical pure states and ubiquity of global minima}
Thermalization in isolated quantum systems shares a key structural principle: the restriction to a small set of quantities of interest, in contrast to the overparameterized total degrees of freedom.     
For instance, the essential aspects of thermalization in isolated quantum systems are the restriction to a limited set of observables and the high dimensionality of the energy shell. In particular, typicality implies that for any fixed observable 
$\hat{O}$ defined on the energy shell ${\cal H}_E$
, the expectation value in a state 
$|\Psi\rangle$, uniformly sampled from the energy shell 
 according to the Haar measure, typically agrees with the microcanonical average. 
More precisely, the probability that the expectation value of a uniformly sampled state deviates from the microcanonical average $\langle\hat{O}\rangle_{mc}$ by more than $\epsilon$ is bounded by the following inequality\cite{Reimann1,Sugita1}. 
\begin{align}
&P(|\langle\Psi|\hat{O}|\Psi\rangle-\langle\hat{O}_{mc}\rangle|\geq\epsilon)\leq\frac{{\rm Var}[\hat{O}]_{mc}}{\epsilon^2(d+1)}, \label{typicality1}
\end{align}
where $d={\rm dim}{\cal H}_E$ denotes the dimension of the energy shell and ${\rm Var}[\hat{O}]_{mc}$ denotes the microcanonical variance.  

Therefore, the vast majority of pure states can be regarded as being in equilibrium\cite{footnote1}. 
From the perspective of state distinguishability, the reduced density matrix of a pure state, given as a mixed state, becomes indistinguishable from the microcanonical ensemble when only observables of a small subsystem are considered\cite{Lebowitz1}.
Here, the total system is divided into subsystems A and B, with dimensions $d_A$ and $d_B$, respectively. The reduced density matrix of subsystem A, $\hat{\rho}_A={\rm Tr}_B|\Psi\rangle\langle\Psi|$, is obtained by tracing out subsystem B. 
In terms of the coefficient matrix of $|\Psi\rangle$ in a product basis $|\Psi\rangle=\sum_{i=1}^{d_A}\sum_{j=1}^{d_B}c_{ij}|\phi_i\rangle_A|\psi_j\rangle_B$, it can be regarded as a normalized  Wishart matrix. 
One motivation for considering the reduced density matrix is that this point corresponds to Wishart-type structures of NTK kernel.

To explore this issue further, let us evaluate the R\'enyi entanglement entropy $S_\alpha=\frac{1}{1-\alpha}\log{\rm Tr}_A\left(\hat{\rho}_A^\alpha\right)$
 of an eigenvector of the Gaussian Unitary Ensemble\cite{Zyckowski1}. 
For large enough subsystems, the R\'enyi entanglement entropy is asymptotically given as 
\begin{align}
&S_\alpha=\log d_A+\frac{1}{1-\alpha}\log\left(\frac{\langle\lambda^\alpha\rangle_{MP}}{\langle\lambda\rangle_{MP}^\alpha}\right), \label{entanglement1}
\end{align}
where $\langle\cdot\rangle_{MP}$ denotes the average with respect to the Mar\v{c}enko-Pastur distribution $\rho_{MP}(\lambda)=\frac{1}{2\pi c\lambda}\sqrt{(\lambda_+-\lambda)(\lambda-\lambda_-)}$, with the ratio $c=\frac{d_A}{d_B}$ and the cutoff $\lambda_\pm=(1\pm\sqrt{c})^2$ in the overparameterized regime\cite{Marcenko1,Bai1,Cheng1}. 
It can be shown that Eq. (\ref{entanglement1}) accurately reproduces the exact values of the purity and the von Neumann entropy for $\alpha=2$ and in the limit $\alpha\rightarrow 1$. Specifically, the purity is given by ${\rm Tr}_A\hat{\rho}_A^2 = \frac{d_A + d_B}{d_A d_B + 1}$\cite{Lubkin1} and asymptotically approaches the value given by (\ref{entanglement1}),  
$\frac{d_A + d_B}{d_A d_B}$, in the limit of large $d_A$ and $d_B$. Similarly, the von Neumann entropy is well approximated by 
\begin{align}
&S\cong
\begin{cases}
\log d_A-\frac{d_A}{2d_B} & (d_A\leq d_B) \nonumber \\
\log d_B-\frac{d_B}{2d_A} & (d_A\geq d_B)   
\end{cases} \nonumber \\
\label{page1-1}
\end{align}
in this limit\cite{Page1}. 
Both the asymptotic evaluation (\ref{entanglement1}) and the exact expressions indicate that the reduced state $\hat{\rho}_A$ is effectively indistinguishable from the microcanonical ensemble $\hat{\rho}_{mc} = \frac{\mathbb{I}_A}{d_A}$ with respect to bipartite entanglement for $d_A\ll d_B$.
Hence, the deviation of the reduced state $\hat{\rho}_A$ from the microcanonical ensemble $\hat{\rho}_{mc}$ is exponentially small due to the overparameterization of the dimension $d_Ad_B$ both for the expectation values and the entanglement measures.

In deep learning, the phenomenon of overparameterization, where the number of parameters far exceeds the number of training data points, has been empirically observed to enhance performance.
The NTK framework provides a theoretical foundation for understanding why overparameterization can lead to successful function approximation, especially in the infinite-width limit\cite{Jakot1}.

Let 
$x_i\in\mathbb{R}^{n_0}$ ($1\leq i\leq N$) denote $N$ input vectors, and let $\theta\in\mathbb{R}^P$
 represent the collection of all trainable parameters of the neural network.
The network defines a function $f(x,\theta)\in\mathbb{R}^{n_L}$ 
that maps input 
$x$ to an output in the $L$-th layer, via a composition of affine transformations and nonlinear activation functions.

Consider a fully connected ANN with parameters 
$\theta\in\mathbb{R}^P$, and output function 
$f(x,\theta)$.
In the NTK regime, one fixes the architecture and defines the Neural Tangent Kernel as:
\begin{align}
&K(x,x')=\sum_{p=1}^P\frac{\partial f(x,\theta)}{\partial\theta_p}\otimes\frac{\partial f(x',\theta)}{\partial\theta_p}, \label{kernel1}
\end{align}
where the derivatives are evaluated at initialization 
$\theta(0)$. 
As the width of each layer tends to infinity, the kernel 
$K(x,x')$ stochastically converges to a deterministic limit, and the function 
$f(x,\theta(t))$ 
 evolves under training in a nearly linear fashion.
 
Since the model is heavily overparameterized, gradient descent modifies the parameters only slightly from their initial values.
This justifies the first-order Taylor expansion of the function 
\begin{align}
&f(x,\theta(t))\cong f(x,\theta(0))+\nabla_\theta f(x,\theta(0))\cdot\left(\theta(t)-\theta(0)\right). \label{linearization1}
\end{align}
Thus, the function evolves approximately linearly in parameter space during training. 

In this regime, training dynamics under gradient descent are effectively governed by kernel regression with the NTK.
For large number of parameters, the cost functional can be minimized to arbitrarily small values, since the function space spanned by the kernel is rich enough to interpolate the training data.

Theoretically, this linearization remains valid under certain Lipschitz continuity conditions\cite{Jakot1}.

Since the parameter changes are small, one can theoretically achieve a sufficiently low cost in the learning dynamics by performing linear regression of the target function using the kernel fixed at its initial value. As a result, a parameter configuration $\theta$ corresponding to a global minimum with near-zero training error exists in the neighborhood of any random initialization. 
Since the cost-minimizing function is entirely determined by the parameter 
$\theta$, we are effectively referring to it as the global minima.

The ubiquity of the global minima can be attributed to the restriction to a small set of observables, namely, $f(x_k,\theta)$ ($1\leq k\leq N$), in the overparameterized regime  
and this shares a structural correspondence with the typicality of pure thermal states\cite{footnote1}. 

Table \ref{correspondence1} summarizes the correspondence between thermalization in isolated quantum systems and function estimation in wide ANN. 
This correspondence is the first main result of this paper. 

\section{Distinguishability and double descent}
To strengthen the correspondence, we examine the counterpart of the double descent phenomenon in thermalization of isolated quantum systems. 
In previous work\cite{blackhole1}, the Page curve (\ref{page1-1})\cite{Page1} for the entanglement entropy $S$ in the black hole evaporation was shown to correspond directly to the double descent phenomenon in overfitting.
Here, we show that this correspondence is not limited to the specific, nontrivial astrophysical phenomenon of black hole evaporation, but holds more generally through the interchange of fixed quantities and variables as shown in the following observation. 

First, we point out that the roles of fixed quantities and variables are interchanged in the Page curve(\ref{page1-1}) and double descent phenomenon. 
Since we consider an isolated quantum system, the total dimension $d_A d_B$ is fixed, while the number of linearly independent observables on subsystem A varies as $d_A$ changes.   
The total dimension of the energy shell $d_A d_B$ corresponds to the number of parameters in ANN.  
Also, the dimension of the subsystem $d_A$ determines the number of linearly independent observables on subsystem A, which is $d_A^2$. 
In the case of double descent, the number of inputs $N$ is kept constant while the number of model parameters is systematically varied.  
Note that the entanglement entropy $S$ attains its maximum at $d_A=d_B$, which corresponds to the fitting threshold in ANN. 
With this point in mind, we now turn to an examination of the Page curve and double descent.

As the dimension $d_A$ increases, the reduced density matrices on subsystem A become more distinguishable, since the number of linearly independent observables on A also increases.
Once $d_A$ reaches the dimension of the energy shell ${\rm dim}{\cal H}_E$, all pure states in ${\cal H}_E$ become perfectly distinguishable, because the density matrix is uniquely specified by the expectation values of all linearly independent observables.
As $d_A$ increases from unity to the square root of ${\rm dim}{\cal H}_E$, the entanglement entropy of the reduced state 
$\hat{\rho}_A$
 increases monotonically, but remains below that of the microcanonical ensemble 
$\hat{\rho}_{mc}$, reflecting the distinguishability of pure states. The reduced state $\hat{\rho}_A$ substantially deviates from the microcanonical ensemble as $d_A$ approaches to $d_B$.     
In particular, the entanglement entropy of a typical state approximately reaches its maximum when 
$d_A=d_B$, and decreases as $d_A$ increases further, owing to the symmetry of entanglement entropy of subsystems A and B. 

Let us consider the counterpart of the maximum entropy condition in ANN. 
It is often emphasized that increasing the number of parameters can lead to overfitting, where the model ends up generalizing a specific function. 
As the number of parameters approaches that of the training data, the generalization of the function $f(x,\theta)$ tends to become unstable.

We provide a quantitative explanation of this scenario based on random matrix theory.
At initialization, the NTK kernel $K(x,x')$ (\ref{kernel1}) can be viewed as a Wishart matrices with non-i.i.d. entries. 
We represent the NTK kernel as an $n_L N\times n_L N$ matrix whose elements are given by $K_{i x_k,i'x_{k'}}$ ($1\leq i, i'\leq n_L$, $1\leq k, k'\leq N$). 
Accordingly, the kernel can be expressed in the form as $K=J J^{\rm T}$, where $J_{ik,p}=\frac{\partial}{\partial\theta_p}f_i(x_k,\theta)\in\mathbb{R}^{(n_L N)\times P}$. 
The fitting threshold is characterized by the condition that the smallest eigenvalue of the kernel approaches zero, causing the inverse to become highly sensitive to small perturbations.
This characterization of the fitting threshold — via the vanishing eigenvalues of the kernel — structurally corresponds to the emergence of zero eigenvalues associated with the exchange symmetry of entanglement entropy at $d_A=d_B$ in isolated quantum systems. 

Formally, the matrix $J$ has dimensions $Nn_L\times P$. 
On the other hand, the total number of parameters $P$ should be effectively reduced to the contributions from the $L$-th layer $P_{\rm eff}$
, since the function $f(x,\theta)$ is a nested composition of affine transformation with asymptotically vanishing scaling factors $\frac{1}{\sqrt{n_l}}$ ($l=0,1,...,L$) and nonlinear activation functions so that the derivative of $f(x,\theta)$ with respect to the parameters of the $l$-th layer ($l\leq L-1$) is negligible. 

Although the entries of 
$K=J J^{\rm T}$  
 are not independent and identically distributed, the smallest eigenvalue of its spectrum follows the same scaling as in the Mar\v{c}enko-Pastur law\cite{Marcenko1,Bai1},
up to an overall constant factor $\sigma$: $\lambda_-\cong\sigma(1-\sqrt{\frac{n_L N}{P_{\rm eff}}})^2$ in the limit of large $n_L N$ and $P$\cite{Vershynin1}.  

Therefore, the fitting threshold occurs when the effective number of parameters $P_{\rm eff}$ equals to the number of scalar outputs, 
i.e., $P_{\rm eff}=n_LN$.
  
Overparameterization refers to the regime where the number of observables, i.e.,  training data points is fixed while the width and therefore the number of parameters is taken to infinity, such that 
$n_L\gg N$. 
When the number of parameters is sufficiently large, small changes across many parameters can accumulate to produce a finite change in the function 
$f(x,\theta)$. In this regime, $f(x,\theta)$ exhibits a high degree of degeneracy, as many different values of 
$\theta$ can yield the same function output. 
This regime, characterized by a number of observables much smaller than 
${\rm dim}{\cal H}_E$, is where typicality emerges. Hence, a direct correspondence with the double descent phenomenon can be established.
Our second main result is the correspondence between the system-size dependence of the entanglement entropy of typical pure states in terms of state distinguishability and double descent phenomenon through the interchange of fixed quantities and variables. And, the condition for the indistinguishability of the reduced density matrices of typical pure states $d_A\ll d_B$ corresponds to the fitting threshold.   
\section{Summary}
We have demonstrated a direct correspondence between thermalization in isolated quantum systems and wide artificial neural networks, based on the notions of overparameterization and the restriction to a limited set of observables.
The typicality of pure thermal states corresponds to the ubiquity of the global minimum in the NTK framework. 
The restriction to a limited class of observables, an essential aspect of thermalization in isolated quantum systems corresponds to the restriction to a single function evaluated only at the input data points, in NTK. 
Furthermore, we investigated the structural correspondence between the roles of Wishart-type matrices in the reduced density matrix of typical pure states and NTK kernel.  
We pointed out that the roles of fixed parameters and variables are reversed between the Page curve and the double descent phenomenon. 
Based on this, we demonstrated that the system size dependence of the entanglement entropy, interpreted in terms of state discrimination, corresponds to the double descent phenomenon.
\begin{acknowledgments}
This work was partly supported by the Grant-in-Aid for Scientific Research (C) (
No.~22K03456) from the Japan Society for the Promotion of Science (JSPS).
\end{acknowledgments}



\begin{thebibliography}{30}
\bibitem{footnote1}A state $|\Psi\rangle$ is being in equilibrium for expectation values of a set of observables, since Eq. (\ref{typicality1}) can simultaneously apply to a set of linearly independent fixed observables $\{{\cal O}_k\}_{k=1}^N$ with $N\ll d$. Here, different observables correspond to a choice of input variable $x_k$.
\end{thebibliography}
\end{document}